\documentclass{mscs}
\usepackage{amssymb, amsmath, epsfig,macros,stmaryrd,xspace}

\begin{document}

\title{Quantum Weakest Preconditions} 
\author[E. D'Hondt and P. Panangaden] 
{E\ls L\ls L\ls I\ls E \ns D\ls '\ls H\ls O\ls N\ls D\ls T $^1$
 \ns and\ns P\ls R\ls A\ls K\ls A\ls S\ls H \ns P\ls A\ls N\ls A\ls N\ls G\ls A\ls 
D\ls E\ls N$^2$\thanks{Ellie D'Hondt  was
funded by the FWO and the VUB (Flanders) and Prakash Panangaden was funded
in part by a grant from NSERC (Canada) and in part by a visiting fellowship
from EPSRC (U.K.).}\\ 
$^1$ Vrije Universiteit Brussel, Belgium
\addressbreak $^2$ McGill University, Montreal, Canada} 
\maketitle

\begin{abstract} 
  We develop a notion of predicate transformer and, in particular, the
  weakest precondition, appropriate for quantum computation.  We show that
  there is a Stone-type duality between the usual state-transformer
  semantics and the weakest precondition semantics.  Rather than trying to
  reduce quantum computation to probabilistic programming we develop a
  notion that is directly taken from concepts used in quantum computation.
  The proof that weakest preconditions exist for completely positive maps
  follows immediately from the Kraus representation theorem.  As an example
  we give the semantics of Selinger's language in terms of our weakest
  preconditions.  We also cover some specific situations and exhibit an
  interesting link with stabilizers.
\end{abstract}

\section{Introduction}

Quantum computation is rapidly becoming a significant topic in theoretical
computer science.  To be sure, there still are essential technological and
conceptual problems to overcome in building functional quantum computers.
Nevertheless there are fundamental new insights into quantum
computability~\cite{Deutsch85,Deutsch92}, quantum
algorithms~\cite{Grover96,Shor94} and into the nature of quantum mechanics
itself~\cite[Part III]{Peres95}, particularly with the emergence of quantum
information theory~\cite[Ch. 12]{Nielsen00}.

These developments inspire one to consider the problems of programming
general-purpose quantum computers.  Much of the theoretical research is
aimed at using the new tools available -- superposition, entanglement and
linearity -- for algorithmic efficiency.  However quantum algorithms are
currently programmed at a very low level -- comparable to classical
computing 60 years ago.  In the search for structure in the space of
quantum algorithms one is led to consider issues like compositionality,
semantics, type systems and logics; these are issues that usually arise in
the context of programming languages.  The present paper is situated in
the nascent area of quantum programming methodology and the design and
semantics of quantum programming languages.  We extend the well-known
paradigm of \emph{weakest preconditions}~\cite{Hoare69,Dijkstra76} to the
quantum context.  The influence of Dijkstra's work on weakest preconditions
has been deep and pervasive and even led to textbook level expositions of
the subject~\cite{Gries81}.  The main point is that it leads to a
\emph{goal-directed} program or algorithm development strategy.  Hitherto
quantum algorithms have been invented by brilliant new insights.  As more
and more algorithms accumulate and a stock of techniques start to
accumulate there will be need for a systematic program development
strategy.  It is this that we hope will eventually come out of the present
work.

In this paper we make two contributions: first, we develop the
appropriate quantum analogue of weakest preconditions and develop the
duality theory.  Rather than reducing quantum computation to
probabilistic computation and using well-known ideas from this
setting~\cite{Kozen81,Kozen85}, we define quantum weakest preconditions
directly.  It turns out that the same beautiful duality between
state-transformer (forwards) and predicate-transformer (backwards)
semantics that one finds in the traditional~\cite{Smyth83,Plotkin83}
and the probabilistic settings~\cite{Kozen85} appears in the quantum
setting.  This is related to the fact that when state transformers are
specified to be completely positive maps, we can prove the existence of
corresponding weakest preconditions in a very general way using a
powerful mathematical result called the Kraus representation
theorem~\cite[Sec. 8.2.4]{Nielsen00}.  In fact the correspondence is very
much more direct in this case than in the case of conventional or
probabilistic languages.

Second, we write the detailed weakest precondition semantics for a
particular quantum programming language.  Quantum programming languages
have started to appear recently. Perhaps the best known is the quantum flow
chart language~\cite{Selinger03}, also referred to as QPL, which is based
on the slogan ``quantum data and classical control''. QPL has a clean
denotational semantics and a clear conceptual basis; we give an alternative
weakest precondition semantics for this language.  It should be noted,
however, that our notion of weakest preconditions and the basic existence
results are \emph{language independent}.

The structure of this paper is as follows.  In \sec{sec2} the
general setup, in particular quantum state transformers and quantum
predicates, is laid out.  Next, in \sec{sec3} we define quantum
weakest preconditions and healthy predicate transformers, proving their
existence for arbitrary completely positive maps and observables.  In
\sec{sec4} we summarize the basic structure of Selinger's
language, and develop its weakest precondition semantics.  We apply our
results to specific situations such as Grover's algorithm and
stabilizers in \sec{sec7}, and conclude with \sec{sec8}.

\section{The quantum framework}\label{sec2}

In this section we define the main concepts on which our theory of quantum weakest preconditions is based. We first give a general overview, after which we specify concrete definitions for quantum states and state transformers in \sec{krtm} and for quantum predicates in \sec{qpreds}. 

Traditionally, there are several means of developing formal semantics for
programming languages.  In the operational semantics for an imperative
language one has a notion of \emph{states}, typically denoted $s$, such
that the commands in the language are interpreted as state transformers.
If the language is deterministic the state transformation is given by a
function, and composition of commands corresponds to functional
composition.  The flow is forwards through the program.  This type of
semantics is intended to give meaning to programs that have already been
written.  It is useful for guiding implementations of programming languages
but is, perhaps, less useful for program development.  By contrast, in a
predicate transformer semantics the meaning is constructed by flowing
backwards through the program, starting from the final intended result and
proceeding to determine what must be true of the initial input.  States are
replaced by \emph{predicates} $p$ over the state space, together with a
satisfaction relation $\models$. Language constructs are interpreted
as predicate transformers.  This type of semantics is useful for
goal-directed programming.  Of course the two types of semantics are
intimately related, as they should be! In a sense to be made precise in
\sec{duality} they are \emph{dual} to each other. The situation for deterministic languages can be found in the first column of Table~\ref{table}.

In the world of probabilistic programs one sees the same duality in action, after suitably generalizing the notions of states and predicates.  
\emph{Probability distributions} now play the role of states.  There are,
of course, states as before and, in a particular execution, there is only one
state at every stage.  However, in order to describe all the possible
outcomes (and their relative probabilities) one keeps track of the
probability distribution over the state space and how it changes during
program execution.  What plays the role of predicates?  Kozen has
argued~\cite{Kozen85} that predicates are measurable functions -- or random
variables, to use the probability terminology.  We note that a special case
of random variables are characteristic functions, which are more easily
recognizable as the analogues of predicates; in fact they \emph{are}
predicates.  In a probabilistic setting one has \emph{expectation} value
rather than truth: truth values now lie in $[0,1]$ rather than in
$\{0,1\}$.  Third, the pairing between measurable functions $f$ and
probability distributions $\mu$ is now given by the integral, which is the
probabilistic expression of the expectation value. These measurable functions are to be viewed as \emph{observations}, which may or may not lead to termination. The pairing between $f$ and $\mu$ then expresses the probability with which termination is achieved when observing $f$. For probabilistic languages the second column of Table~\ref{table} summarizes the main concepts.

For the quantum world we again need a notion of state -- or, more precisely, probability
distributions over possible states -- a notion of predicate, and a pairing.
Our choices are very much guided by the probabilistic case, but we are
\emph{not} claiming that quantum computation can be seen as a special case
of classical probabilistic computation.  Instead, we take \emph{density matrices} as
the analogue of probability distributions, while for predicates we take the
\emph{observables} of the system.  These are given by (a certain restricted
class of) Hermitian operators.  Finally, the notion of a pairing is again
the expectation value, but given by the rules of quantum mechanics; that is
we have $\tr{M\rho}$, where tr stands for the usual trace from linear
algebra, $\rho$ is a density matrix and $M$ an observable.  Throughout this
paper we work with finite-dimensional Hilbert spaces and one can think of
$M$ and $\rho$ as matrices.  We discuss these concepts in more depth in \ses{krtm}{qpreds};
a summary can be found in the last column of Table~\ref{table}. Note however that, just as for the probabilistic case, the pairing $\tr{M\rho}$ may be interpreted as the probability of termination when observation $M$ is made in the state $\rho$.

\begin{table}
\caption{Comparing situations.}
\begin{tabular}{|c|c|c|}
\hline Deterministic  & Probabilistic & Quantum \\
\hline
states & probability distributions & density matrices\\
$s$& $\mu$ & $\rho$\\
\hline
predicates & measurable functions & observables \\
$p$ & $f$ & $M$ \\
\hline
satisfaction & expectation value & quantum expectation value \\
$s \models p$ & $\int f d\mu$ & $\tr{M \rho}$ \\
\hline
\end{tabular}
\label{table}
\end{table}%

Why cannot one just use probabilistic predicates and the general theory of probabilistic predicate
transformers in a quantum context?  The following simple example -- due to one of the referees --
illustrates why.  Suppose that we have a two-dimensional Hilbert space of
states with basis vectors written $\0$ and $\1$.  Two other
states in this Hilbert space are $\frac{1}{\sqrt{2}}(\0+\1)$ and
$\frac{1}{\sqrt{2}}(\0-\1)$.  We use the notation $\ens{\gket}$ for the density matrix
$\proj{\psi}$ and write convex combinations like $\la\ens{\gket} +
(1-\la)\ens{\gket}$ for the density matrix of a mixed state, i.e. an ensemble.  Now consider the measurable function $f$ defined by:

\EQ{\SP{
 f(\0)  &=  0\\
 f(\1)  &=  0\\
 f(\frac{1}{\sqrt{2}}(\0+\1))  &=  1\\
 f(\frac{1}{\sqrt{2}}(\0-\1)  &=  1\text{  .}
}}
This function is indeed measurable but not linear and cannot correspond to
any kind of physical observable or measurement.  To see what happens,
consider the ensemble $\rho = \frac{1}{2}\ens{\0} +
\frac{1}{2}\ens{\1}$.  When $f$ is applied to this one obtains $0$.
However, when $f$ is applied to the ensemble 
$\rho' =
\frac{1}{2}\ens{\frac{1}{\sqrt{2}}(\0+\1)} + 
\frac{1}{2}\ens{\frac{1}{\sqrt{2}}(\0-\1)} $ 
we obtain the value $1$.  The point is that $\rho$ and $\rho'$ are physically
indistinguishable, and thus one cannot have a physical observable that tells
these ``two'' ensembles apart.  When developing a theory of predicates and
predicate transformers one must therefore restrict to mathematical objects
that are compatible with the linear structure of quantum mechanics.  It is
a conceptual error to think that quantum mechanics can be understood just
with probabilistic constructs. We note that the work in~\cite{Butler99}, which uses probabilistic predicates to analyze Grover's algorithm~\cite{Grover96}, avoids this conundrum because it considers only pure-state situations.

\subsection{Quantum states and state transformers}\label{krtm}

Typically a quantum system is described by a Hilbert space, physical
observables are described by Hermitian operators on this space and
transformations of the system are effected by unitary
operators~\cite{Peres95}.  However, we need to describe not only so-called
\emph{pure} states but also \emph{mixed} states.  These arise as soon as
one has to deal with partial information in a quantum setting.  For
example, a system may be prepared as a statistical mixture, it may be mixed
as a result of interactions with a noisy environment (decoherence), or by
certain parts of the system being unobservable.  For all these reasons we
need to work with \emph{probability distributions} over the states in a
Hilbert space.  In quantum mechanics this situation is characterized by
\emph{density matrices}, of which a good expository discussion appears in
~\cite[Ch. 2]{Nielsen00}.  Concretely, a density matrix $\rho$ on a
Hilbert space \cH is a \emph{positive operator}, that is, for all states $\ket{x}$ in \cH one requires that $\braket{x}{\rho x}\geq 0$, with furthermore $\mrm{tr}\rho\leq 1$.  The
reason why we do not have the usual equality is that we do not assume that
everything is always normalized.  Hence, in order to interpret a density
matrix as a probability distribution one first needs to renormalize if
necessary.  This is a bit of a nuisance if one wants a direct
interpretation of the density matrix at every stage of the computation;
however, one does recover the probabilities correctly if one starts with a
normalized density matrix at the start of a computation and multiplies out
everything at the end.  This convention saves some notational overhead and
is used by Selinger~\cite{Selinger03}.  We denote the set of all density
matrices over a Hilbert space \cH by $\dm$.

As we have mentioned in the above, forward operational semantics is described by quantum state
transformers.  The properties of such state transformers are now well
understood.  A physical transformation must take a density matrix to a
density matrix.  Thus it seems reasonable to require that physical
operations correspond to positive maps, which are linear maps that take a
positive operator to a positive operator.  However, it is possible for a
positive map to be tensored with another positive map - even an identity
map - and for the result to fail to be positive.  Physically this is a
disaster.  Indeed, this means that if we formally regard some system as part
of another far away system which we do not touch (that is, to which we apply
the identity transformation), then suddenly we have an unphysical
transformation.  A simple example is provided by the transpose operation, which
is a positive map while its tensor with an identity is not.
Therefore, we need the stronger requirement that physical operations are
\emph{completely} positive, a property which is defined as follows.

\DEF{
A map \cE is \emph{completely positive} when it takes density matrices to
density matrices, and likewise for all trivial extensions $I\ox \cE$.  
\label{defA2}
}
Note that such a map may operate between distinct Hilbert spaces, that is
in general we have $\cE:\den{\cH_{1}}\rar \den{\cH_{2}}$.  We denote by
$\cpt$ the set of all such maps, and write $\cp$ for $\cpv{\cH}{\cH}$.  

We frequently rely on the Kraus representation theorem for completely
positive maps.   

\THM{[Kraus Theorem]
The map $\cE:\den{\cH_{1}}\rar \den{\cH_{2}}$ is a completely positive map
if and only if for all $\rho \in  \den{\cH_{1}}$ we have that 
\EQ{
 \cE(\rho) = \sum_{i}E_{i}\rho E_{i}^{\dag}
\label{eqA5}
}
for some set of operators $\ens{E_{i}:\cH_{1}\rar \cH_{2}}$, with
$\sum_{i}E_{i}^{\dag}E_{i} \leq I$.   
\label{kraus}
}
The condition on the $E_i$ ensures that trace of the density matrix
never increases.  \eq{eqA5} is also known as the
\emph{operator-sum representation}.  The proof to this theorem can be
found, for example, in \cite[Sec. 8.2.4]{Nielsen00}.  Note there is
nothing in the theorem that says that the $E_i$ are unique.

\subsection{Quantum predicates}\label{qpreds}

In this section, we define quantum predicates and the
associated order structure required for the development of our theory.
Concretely, we need an ordering on predicates so as to define
\emph{weakest} preconditions, and this order should be
\emph{Scott-continuous} in order to deal with programming language aspects
such as recursion and iteration.

As argued above, quantum predicates are given by Hermitian operators.
However, general Hermitian operators will not yield a satisfactory logical
theory with the duality that we are looking for.  We need to restrict to
positive operators and - in order to obtain least upper bounds for
increasing sequences - we need to bound them.  More precisely, we have the
following definition.

\DEF{
  A \emph{predicate} is a positive - hence Hermitian - operator with the
  maximum eigenvalue bounded by $1$.
\label{def2.22}
}
The reason for taking predicates to have the maximum eigenvalue bounded by 1 
is in order to get a complete partial order (CPO); we clarify
this below.  Since our predicates are positive operators their eigenvalues
are real and positive.  We denote the set of all predicates on a
Hilbert space $\cH$ by $\pred$. 

\PRO{\label{predpos}
For any density matrix $\rho$ and Hermitian operator $M$ we have 
$0\leq \tr{M\rho}\leq 1$ if and only if $M$ is positive and its eigenvalues
are bounded by $1$.
} 
\begin{proof}
Note that for any element $\ket{\psi}$ of $\cH$ we have
$\tr{M\proj{\psi}} = \mel{\psi}{M}{\psi}$.  Assume that 
  $0\leq\tr{M\rho}\leq 1$ for any $\rho$ a density matrix.  Choose $\rho=
  \proj{\psi}$ where $\ket{\psi}$ is an arbitrary normalized vector.  We
  have $0 \leq \tr{M \proj{\psi}} = \mel{\psi}{M}{\psi}$, which says that
  $M$ is positive.  Now choose $\ket{\psi}$ to be a normalised eigenvector
  of $M$ with eigenvalue $\lambda$, necessarily real and positive, so we
  have that $\tr{M \proj{\psi}} = \mel{\psi}{M}{\psi} =
  \lambda\braket{\psi}{\psi} = \lambda \leq 1$.  Thus the eigenvalues are
  bounded by $1$.  The converse is obvious once we note that any density
  matrix is a convex combination of density matrices of the form
  $\proj{\psi}$.
\end{proof}

Thus we could have defined predicates as positive operators $M$ such that
for every density matrix $\rho$ we have $0\leq \tr{M\rho}\leq 1$.  This
exhibits the predicates as ``dual'' to density matrices.

We define an ordering as follows.
\DEF{ 
For matrices $M$ and $N$ in $\comp^{n \x n}$ we define $M \prec N$ if
$N-M$ is positive.  
\label{def2.23}
} 
This order is known in the literature as the \emph{L\"owner partial
order} \cite{Lowner34}.  Note that this definition can be rephrased in
the following way, where $\dm$ denotes the set of all density matrices.

\PRO{
$M \prec N$ if and only if $\forall \rho \in \dm .  \tr{M\rho}\leq
\tr{N\rho}$   
\label{eq2.12}
\label{prop2.21}
}
\begin{proof}
Indeed, $N-M$ positive means that for all $x \in \cH$ we have
$\braket 
{x}{N-M|x} \geq 0$, or, equivalently, $ \tr{ (N-M).  \proj{x}{x}}
\geq 0$.  By linearity of the trace and the fact that the spectral
theorem holds for all $\rho \in \dm$ we obtain the desired result. For the converse,
take all pure states $\rho=\proj{x}$.  Then we find that for
all $x \in \cH$ we have $\braket{x}{N-M|x} \geq 0$, or in other words $M
\prec N$.  
\end{proof}

Put otherwise, $M \prec N$ if and only if the expectation value of $N$
exceeds that of $M$.  With the above definitions, we have the following
result.

\PRO{
The poset $(\pred, \prec)$ is a complete partial order (CPO),
i.e.  it contains least upper bounds of increasing sequences.   
\label{prop2.23}
}

\begin{proof}
Take an increasing sequence of predicates $M_{1}\prec M_{2}\prec \ldots
\prec M_{i}\prec \ldots$.  This is a sequence of positive operators with trace bounded
by 1, or in other words, density matrices.  Since $(\dm,\prec)$ is a
CPO \cite{Selinger03}, this sequence has a least upper bound $M$.  
It follows that $(\pred, \prec)$ is a CPO.  
\end{proof}
Taking predicates to be \emph{bounded} Hermitian operators leads to
\pro{prop2.23}, which guarantees the existence of fixpoints and
thus allows for the formal treatment of iteration and recursion in
\sec{sec4}.

\section{Quantum weakest preconditions and duality} \label{sec3}
In this section we elaborate our theory of quantum weakest preconditions. We first give the main definitions in \sec{defs}, after which we explore healthiness conditions in \sec{healthiness}. Next, we investigate weakest precondition predicate transformers for completely positive maps in \sec{wprforcpmaps}. With the latter results we obtain a duality between the forward state transformer semantics and the backward weakest precondition semantics in \sec{duality}.

\subsection{Definitions}\label{defs}

In a quantum setting, the role of the satisfaction relation is taken
over by the \emph{expectation value} of an observable $M$, just as for
probabilistic computation.  The quantum expectation value of a predicate
$M$ is given by the trace expression $\tr{M\rho}$.  Preconditions for
a quantum program $\cQ$ -- described in an unspecified quantum
programming language -- are defined as follows.  We write $\cQ$ for the
program as well as for the trace-nonincreasing completely positive map that
it denotes.

\DEF{\label{prec}
The predicate $M$ is said to be a \emph{precondition} for the predicate
$N$ with respect to a quantum program $\cQ$, denoted  $M \{\cQ\} N$, if
\EQ{
\forall \rho \in \dm .  \tr{M\rho} \leq \tr{N \cQ(\rho)}
\label{eq3.0}
}
\label{def3.1}
}
We also introduce the  notation $\rho\models_r M$ to mean that $\tr{M\rho}\geq r$. Thus we think of
this as a quantitative satisfaction relation with the real number $r$
providing a ``threshold'' above which we deem that $\rho$ satisfies
$M$.  

The exact syntax of the quantum program $\cQ$ is left unspecified
deliberately, as we want to state these definitions without committing to
any particular framework. Of course we expect \cQ to implement at least
some transformation on density matrices, in particular we may think of \cQ
as implementing a completely positive map. Note however, that \de{prec}, as
well as \de{def3.2} below, does not exclude other possibilities. For
example we could also investigate possibilities proposed in~\cite{Shaji05},
where it is argued that positive but not completely positive or even not
positive maps are also good candidates for describing open quantum
evolutions.

This definition deserves motivation.  If all density matrices were
normalized then it is easy to motivate \de{prec}: if we want the expectation
value of $N$ in the state $\cQ(\rho)$ to be above some real number
$r$, say, then this is guaranteed if the expectation value of $M$ in
the state $\rho$ is above $r$.  In the case of our unnormalized
density matrices we have to do a little calculation to see that the same
holds.  We write the expectation value of $M$ in a state (density matrix)
$\rho$ as \expv{M}{\rho}.  Now we assume that $M,N$ and $\cQ$ satisfy the
conditions of \de{prec}.  Let $\rho$ be any (unnormalized) density matrix
and let its normalized version be $\overline{\rho} = \rho/\tr{\rho}$.  Then
we have 
\EQ{\SP{
\expv{M}{\rho}& =    \tr{M\overline{\rho}}\\
& =    \frac{1}{\tr{\rho}}\cdot \tr{M\rho}\\
& \leq \frac{1}{\tr{\rho}}\cdot \tr{N\cQ(\rho)}\\
& =    \frac{\tr{\cQ(\rho)}}{\tr{\rho}}\cdot   \frac{1}{\tr{\cQ(\rho)}} \tr{N\cQ(\rho)} \\
& =    \frac{\tr{\cQ(\rho)}}{\tr{\rho}}\cdot \expv{N}{\cQ{\rho}}\\
& \leq \expv{N}{\cQ(\rho)}\text{  .}
}}
Thus, even though the density matrices are not normalized and we cannot
read the expectations \emph{directly} at every intermediate stage,
\de{prec} still has the same import as in the normalized case, as well  as in
the case of probabilistic predicate transformers. 

From this we define weakest preconditions in the usual way.

\DEF{
A \emph{weakest precondition} for a predicate $M$ with respect to a
quantum program $\cQ$, denoted $\wpr{\cQ}(M)$, is such that for all
preconditions $L\{\cQ\} M$ implies  $L \prec \wpr{\cQ}(M)$.   
\label{def3.2}
}
Note that \emph{weakest} in this context is equal to \emph{largest};
indeed, a larger predicate means that \eq{eq3.0} holds for more initial
states $\rho$, and 
thus corresponds to a weaker constraint.  The weakest precondition predicate transformer for a program $\cQ$, if it exists, is denoted $\wpr{\cQ}:\predv{\cH_{2}}\rar\predv{\cH_{1}}$, where $\cH_{2}$ and $\cH_{1}$ are the output and input Hilbert spaces respectively.

\subsection{Healthiness conditions} \label{healthiness}

In analogy with \cite{Dijkstra76}, we want to formulate
\emph{healthiness conditions} for quantum predicate transformers.  These
are important because they characterize exactly those programs that can
be given a weakest precondition semantics which is
dual to its forwards state transformer semantics.  Moreover, healthiness conditions allow one to prove general laws for reasoning about programs.  The healthiness conditions we propose for the quantum case are \emph{linearity} and \emph{complete positivity}, leading to the following definition. 

\DEF{
A \emph{healthy} predicate transformer $\al: \predv{\cH_{2}}\rar\predv{\cH_{1}}$ is a predicate transformer that is \emph{linear} and \emph{completely positive}, i.e. it it takes predicates to
predicates and likewise for all trivial extensions $I\ox \al$.  
We denote the associated space of healthy predicate transformers as $\prtft$.
\label{def2.24}
}

As we shall see in the following section these conditions all hold in the
framework where quantum programs correspond to completely positive maps.
Linearity is certainly a requirement in the inherently linear context of
quantum mechanics, as the example given in \sec{sec2} clearly shows. Just
as in the probabilistic case~\cite{Morgan04}, linearity implies the
analogues of some of the healthiness conditions for deterministic programs,
namely feasibility, which means that $\wpr{\cQ}(0) = 0$, monotonicity and
continuity. These proofs are easy and are left to the reader.  The
requirement that predicate transformers should be completely positive on
$\pred$, is a very natural one.  Indeed, if $\al$ is a predicate
transformer, which acts only on part of a composite Hilbert space $\cH$,
then composing it with the identity predicate transformer working on the
rest of the Hilbert space should still result in a valid predicate
transformer.

We equip $\prtft$ with an order structure by extending the L\"{o}wner
order on predicates in the following way.
\DEF{ 
For healthy predicate transformers $\al$ and $\ba$ in $\prtft$ we define $\al \prec \ba$ if
$\ba-\al$ is a healthy predicate transformer.  
\label{healthy}
} 
If $\al\prec\ba$ then for all predicates $M\in\predv{\cH_{2}}$ we have that $\al(M)\prec\ba(M)$, where $\al(M)$ and $\ba(M)$ are predicates on $\cH_{1}$. Requiring only this would be the obvious extension of the L\"owner order, however, since we are working in the space of healthy predicate transformers we also need to demand that $\ba-\al$ is completely positive. That is,  for all extended predicates $M_{e}\in\predv{\cH_{2}\ox\cH}$ we have $(\al\ox I_{\cH})(M_{e})\prec(\ba\ox I_{\cH})(M_{e})$.  We then have the following result.  

\PRO{
The poset $(\prtft, \prec)$ is a CPO.
\label{prop2.24}
}

\begin{proof}
Take an arbitrary increasing sequence of predicate transformers
\[ \al_{1}\prec \al_{2}\prec \ldots \prec \al_{i}\prec \ldots .\]
This is in fact a sequence of completely positive maps.  Hence since
$(\cpt,\prec)$ is a CPO \cite{Selinger03}, this sequence has a least upper
bound $\al$.  It follows that $(\prtft, \prec)$ is a CPO.
\end{proof} 

Note that the CPO structure as defined on predicates $\pred$ and
associated predicate transformers $\prtf$ is identical to
that for density matrices $\dm$ and associated completely positive maps
$\cp$, as defined in \cite{Selinger03}.

Furthermore, for healthy predicate transformers, we have the following
immediate consequence of Kraus's theorem.

\PRO{
The operator $\al$ is a healthy predicate transformer if and only if
one has that
\EQ{
\forall M \in \pred .  \al(M)=\sum_{u}A_{u}^{\dag} M A_{u}
\label{eq2.22}
}
for some set of linear operators $\ens{A_{u}}$ such that
$\sum_{u}A_{u} A_{u}^{\dag} \leq I$.  
\label{theorem2.1}
}

\subsection{Predicate transformers for completely positive
  maps}\label{wprforcpmaps} 

Let us now consider the following framework: the forward semantics of a
quantum program \cQ is given by a trace-nonincreasing completely positive
map $\cE \in \cpt$, which we write as $\sem{\cQ}=\cE$.  In this section we
prove an existence theorem of weakest preconditions for completely positive
maps, and show that they satisfy the healthiness conditions given in
\sec{healthiness}, i.e.  that they are healthy predicate transformers.

\PRO{
$\forall \cE \in \cpt $ and $N\in \pred$, $\wpr{\cE}(N)$ exists and is
unique.  Furthermore, we have that 
\EQ{
\forall \rho .  \tr{\wpr{\cE}(N)\rho} = \tr{N \eff{\rho}}
\label{eq3.1}
}
\label{prop3.1}
}

\begin{proof}

To prove existence, take an arbitrary predicate $N$ and operation $\cE$.  
From the Kraus representation theorem stated in \sec{krtm}, one
has for every operation $\cE$ that 

\EQ{
\eff{\rho}=\sum_{m}E_{m}\rho E_{m}^{\dag}\label{eq3.11}
}
with $\sum_{m}E_{m}^{\dag} E_{m}\leq I$.  Using this, together with the
fact that the trace is linear and invariant under cyclic
permutations, we obtain for a predicate $N$ that 

\EQ{
\tr{N\eff{\rho}}= \tr{(\sum_{m}E_{m}^{\dag} N E_{m})\rho}
\label{eq3.2}
}
If we then take
\EQ{
M=\sum_{m}E_{m}^{\dag} N E_{m}
\label{eq3.3}
}
in \eq{eq3.2}, we obtain
\EQ{
\forall \rho .  \tr{M\rho} = \tr{N \eff{\rho}}
\label{eq3.4}
}
So $M$ is a precondition for $N$ with respect to $\cE$.  Now take any
other precondition $M'$ for $N$ with respect to $\cE$.  In other words 
\EQ{
\forall \rho .  \tr{M'\rho} \leq \tr{N \eff{\rho}}
\label{eq3.5}
}
but because of \eq{eq3.4} and \pro{prop2.21}, this implies that $M'\prec
M$.  So M is the weakest precondition for $N$ with respect to $\cE$, denoted
$\wpr{\cE}(N)$.

To prove uniqueness, suppose  the predicate $P$ is also a weakest
precondition for $N$ with respect to $\cE$.  Then we have $M\prec P$,
but also, since $M$ is a weakest precondition, $P\prec M$.  But then,
since $\prec$ is an order, we have $M=P$.  
\end{proof}

From \eq{eq3.3} and \pro{theorem2.1} we obtain the following.

\COR{
For all $\cE \in \cp$, $\wpr{\cE} \in \prtf$, i.e.  it is a healthy predicate transformer.
\label{prop3.2}
}

\subsection{Duality}\label{duality}

In this section, we investigate the duality between the forward semantics
of completely positive maps as state transformers, and the backwards
semantics of healthy predicate transformers.  This duality is part of a web
of dualities known to mathematicians as Stone-type
dualities~\cite{Johnstone82}, the prototype of which is the duality between
boolean algebras and certain topological spaces called Stone spaces.  For
readers with a background in category theory we note that such a duality is
captured by an adjoint equivalence mediated by a pairing, for example the
satisfaction relation between states and predicates.  Kozen - following
suggestions of Plotkin - found such a duality in the context of
probabilistic programs~\cite{Kozen85}.  We show that such a duality exists
in the quantum setting as well.

In the quantum context, we find the duality by defining an isomorphism
between the set of all completely positive maps $\cpt$ and the set of all
healthy predicate transformers $\prtft$.  We can associate a healthy
predicate transformer with every operation $\cE\in\cpt$; this follows
immediately from \pro{prop3.1}.  Indeed, we associate with every operation
$\cE$ its weakest precondition predicate transformer $\wpr{\cE}$.  To
complete the duality, we need to associate an operation $\cA \in \cpt$ with
a predicate transformer $\al \in \prtft$.  Using the operator-sum
representation for predicate transformers as given in \eq{eq2.22}, we have
that

\begin{align}
\tr{\al(M) \rho}&= \tr{(\sum_{u}A_{u}^{\dag}MA_{u})\rho} \notag \\
&=\tr{M.(\sum_{u}A_{u}\rho A_{u}^{\dag}}
\label{eq3.23}
\end{align}
If we then take

\EQ{
\cA(\rho)=\sum_{u}A_{u}\rho A_{u}^{\dag}
\label{eq3.24}
}
we obtain 

\EQ{
\tr{\al(M) \rho}= \tr{M\cA(\rho)}
\label{eq3.22}
}
thus associating a state transformer with every healthy predicate
transformer.  Analogously to the above, one could say that this
expression defines the ``strongest post-state'' $\cA(\rho)$ for a state
$\rho$, with respect to a predicate transformer $\al \in \prtf$.

To see this as a duality more clearly, we use the notation
$\rho\models_r M$ defined in \sec{sec3}. Then we have 
\EQ{
\frac{\cE(\rho)\models_r M}{\rho\models_r
\wpr{\cE}{M}}.
} 
It is straightforward to see that we have an order isomorphism between the
domain of predicate transformers $\prtft$ and the domain of state transformers $\cpt$, and this for arbitrary Hilbert spaces $\cH_{1}$ and $\cH_{2}$. As an aside we note that because of this and the fact that maps in   $\prtft$ are Scott-continuous, we immediately obtain that healthy predicate transformers are Scott-continuous as well.

\section{Weakest precondition semantics for QPL} \label{sec4}

The quantum flow chart language or Quantum
Programming Language (QPL), is a typed programming language for quantum
computation with a formal semantics, which is built upon the idea of quantum data and classical control~\cite{Selinger03}.  It is very different from previously
defined quantum programming languages, which do not have a formal
semantics and are imperative rather than functional.  Syntactically, programs in QPL are represented either by flow charts or
by QPL terms.  The basic language constructs are allocating or discarding bits or qubits, assignment,
branching, merge, measurement and unitary transformation.  One can then
build more complex programs from these atomic flow chart components
through context extension, vertical and horizontal composition, iteration and recursion.  

At each moment the denotation of the system, called a \emph{state} in~\cite{Selinger03}, is given by a tuple of density matrices.  The tuple dimension originates from classical bits present in the program, while tuple entries represent the state of all available qubits as density matrices.  Each member of the tuple corresponds to a particular instantiation of the classical variables in lexicographical order; this is otherwise interpreted as a classical control path.  Concretely, a state for a typing context containing $n$ bits and $m$ qubits is given by a $2^{n}$-tuple $(\rho_{0},\dots,\rho_{2^{n}-1})$ of density matrices in $\den{\mbb C^{2m}}$.  Program transformations are given by tuples of trace-decreasing completely positive maps which act on states -- these are called superoperators in~\cite{Selinger03}.  Note that positivity on tuples is defined such that it holds for each entry, while the trace of a tuple is defined as the sum of the traces of its entries.  

The formal semantics of QPL is developed within the category $\bQ$, which has
signatures (which define tuples of complex finite-dimensional vector spaces) as its objects and
superoperators as its morphisms.  This category is equipped with a
CPO-structure, composition, a coproduct $\oa$ and a tensor product $\ox$, all of which are Scott-continuous, and a monoidal trace \tbf{Tr}.  The latter is just the categorical trace for the co-pairing map $\oa$; as per~\cite{Selinger03} we use the term monoidal to avoid confusion with the categorical trace for the tensor product, i.e.  the matrix trace tr.  The coproduct $\oa$ denotes concatenation of signatures.  Note that, unlike the very similar situation of finite-dimensional vector spaces, it is \emph{not} a product, as $\oa$ does not respect matrix traces and hence is not a superoperator.  All basic flow chart components are
morphisms of this category.  For example,the semantics of measurement of one qubit $q$is defined as
\EQ{
\sem{\mrm{measure}\; q}: \mbf{qbit}\rar \mbf{qbit}\oa \mbf{qbit}: \rho \rar (\cE_{0}\oa \cE_{1})(\rho) = P_{0 }\rho P_{0 }\oa P_{1 }\rho P_{1}\text{  ,}
\label{measure}
}
where $P_{\psi}=\proj{\psi}$.
Context extension is modeled by specific $\oa$ or $\ox$ operations on the state.  Vertical and horizontal composition correspond to composition and coproducts of
morphisms respectively, while iteration is interpreted via the monoidal
trace.  Specifically, suppose that an operation $\cE: \sig \oa \tau
\rar  \sig' \oa \tau$, where $\sig$, $\sig'$ and $\tau$ are
signatures, has been decomposed into components $\cE_{11}: \sig
\rar  \sig'$, $\cE_{12}: \sig \rar \tau$, $\cE_{21}:
\tau \rar  \sig'$ and $\cE_{22}:\tau \rar \tau$.  The
operation obtained from $\cE$ by iterating over $\tau$ is then given by
the \emph{monoidal trace of} $\cE$, defined as

\EQ{
\Tr{\cE}=\cE_{11}+\sum_{i=0}^{\infty}\cE_{21};\cE_{22}^{i};\cE_{12}\text{  .}
\label{eq4.1}
}
The existence of this limit is ensured by the CPO structure on superoperators~\cite{Selinger03}.  

QPL also allows recursively defined operations $\cE=F(\cE)$, where $F$ is a
flow chart.  In this case, $F$ defines a Scott-continuous function
$\Phi_{F}$ on morphisms, such that the interpretation of $\cE$ is given
as the least fixed point of $\Phi_{F}$.  Concretely,
\begin{align}
\cE &=\lub{i}F_{i} \qquad \text{with $F_{0}=0$ and $F_{i+1}=\Phi_{F}(F_{i})$}
\label{eq4.2}\\
&= \lub{i}\Phi_{F}^{i}(0)\text{  ,}
\label{eq4.3}\\
\notag
\end{align}
where $0$ is the zero completely positive map, which corresponds to the
divergent program.  Again, the existence of these fixed points is
ensured by the CPO structure.

In what follows we derive a weakest precondition semantics for QPL.  Note that in order to to this, our predicates need to operate on tuples of density matrices.  We do this by writing expressions of the type $M_{1}\oa M_{2}$ where $M_{1}$ and $M_{2}$ are predicates in the sense of \de{def2.22}.  This works since $\oa$ is in fact defined on arbitrary linear maps.  We frequently write $\wpr{\cQ}$
instead of $\wpr{\sem{\cQ}}$; by this we mean that we use the forward semantics of \cQ, which is given by a tuple of  completely positive maps, to derive the weakest precondition predicate transformer for \cQ according to the results in \sec{wprforcpmaps}.

\paragraph{Basic flow charts.} In our approach we uniformly consider all basic flow charts
to be operations in the operator-sum representation
as in \eq{eq3.11}.  As such \pro{prop3.1} already provides
a weakest precondition semantics for these atomic flow charts.  Note, however, that predicates need to be defined in accordance with the type of the tuple exiting a basic flow chart. As a concrete example, we mention measurement, for which the forward semantics is specified in \eq{measure}.  We find that for all predicates $M_{1}\oa M_{2}$ we have 
\EQ{ \SP{
\wpr{\mbf{measure} \; q}(M_{1}\oa M_{2})& = \wpr{\cE_{0}\oa\cE_{1}}(M_{1}\oa M_{2})\\
&= \wpr{\cE_{0}}(M_{1})+\wpr{\cE_{1}}(M_{2})\\
&= P_{0 }M_{1} P_{0 }+ P_{1 }M_{2} P_{1}\text{  .}
\label{measurewp}
}}

We now turn towards weakest precondition relations for composition techniques of QPL.

\begin{figure}

\scalebox{.50}{\includegraphics{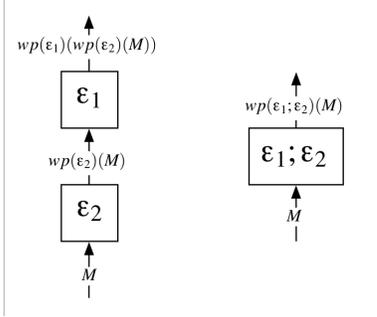}}
\caption{Sequential composition schematically.}

\label{sequential}
\end{figure}

\paragraph{Sequential composition.} Suppose we take the sequential composition of two operations $\cE_{1}$
and $\cE_{2}$, as shown in \fig{sequential}.  For the composed
operation $\cE_{1};\cE_{2}$ and for all predicates $M$ we have that 

\EQ{
\tr{M .  (\cE_{1};\cE_{2})(\rho)}=\tr{\wpr{\cE_{1};\cE_{2}}(M).  \rho}\text{  .}
\label{eq5.11}
}
If we calculate weakest preconditions for both
operations separately and then compose them sequentially, we obtain 

\EQ{\SP{
\tr{M.  (\cE_{1};\cE_{2})(\rho)}&=\tr{M.  \cE_{2}(\cE_{1}(\rho))}\\
&=\tr{\wpr{\cE_{2}}(M).\cE_{1}(\rho)}\\
&=\tr{\wpr{\cE_{1}}(\wpr{\cE_{2}}(M)).\rho}\\
&=\tr{(\wpr{\cE_{2}};\wpr{\cE_{1}})(M).\rho}\text{  .}
\label{eq5.12}
}}
Hence by \eqs{eq5.11}{eq5.12} we obtain that weakest predicate transformers compose sequentially as follows,

\EQ{
\wpr{\cE_{1};\cE_{2}}=\wpr{\cE_{2}};\wpr{\cE_{1}}\text{  .}
\label{eq5.13}
}
This is the same rule as one finds for sequential composition in classical programming languages~\cite{Dijkstra76}.

\begin{figure}

\scalebox{.50}{\includegraphics{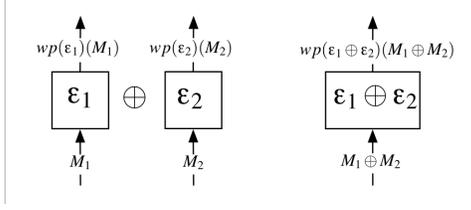}}
\caption{Parallel composition schematically.}

\label{parallel}
\end{figure}

\paragraph{Parallel composition.} Suppose we take the parallel composition of two operations $\cE_{1}$
and $\cE_{2}$, as shown in \fig{parallel}.  For the composed
operation $\cE_{1}\oa \cE_{2}$  we have that 
\EQ{
\tr{(M_{1}\oa M_{2}) .  (\cE_{1}\oa\cE_{2})(\rho_{1}\oa
  \rho_{2})}=\tr{\wpr{\cE_{1}\oa\cE_{2}}(M_{1}\oa M_{2}) .  (\rho_{1}\oa
  \rho_{2})} \text{  .}
\label{eq5.21}
}
On the other hand, if we calculate weakest preconditions for both
operations separately and then compose them in a parallel way, we
obtain 
\EQ{\SP{
\tr{(M_{1}\oa M_{2}) .  (\cE_{1}\oa\cE_{2})(\rho_{1}\oa \rho_{2})}
&=\tr{M_{1} .  \cE_{1}(\rho_{1})\oa M_{2} .  \cE_{2}(\rho_{2})}\\
&=\tr{M_{1} .  \cE_{1}(\rho_{1})}+\tr{M_{2} .  \cE_{2}(\rho_{2})}\\
&=\tr{\wpr{\cE_{1}}(M_{1}) .  \rho_{1}}+\tr{\wpr{\cE_{2}}(M_{2}) .  \rho_{2}}\\
&=\tr{(\wpr{\cE_{1}}(M_{1}) \oa \wpr{\cE_{2}}(M_{2})) .  (\rho_{1}\oa\rho_{2})}\\
&=\tr{(\wpr{\cE_{1}}\oa\wpr{\cE_{2}})(M_{1}\oa M_{2}).(\rho_{1}\oa \rho_{2})} \text{  .}
\label{eq5.22}
}}
Comparing \eqs{eq5.21}{eq5.22} we obtain that for
parallel composition weakest precondition predicate transformers
compose as follows,
\EQ{
\wpr{\cE_{1}\oa\cE_{2}}=\wpr{\cE_{1}}\oa\wpr{\cE_{2}}
\label{eq5.23}
}

\paragraph{Context extension.} Let us now study what occurs if we weaken a context with dummy
classical or quantum variables.  Suppose first that we have a QPL program \cQ with denotation \cE.  We first modify \cQ by picking a fresh classical variable $b$ and adding it to \cQ's context; denote the resulting program $\cQ_{b}$.  The forward semantics of the latter is given by $\cE\oa\cE$ ~\cite{Selinger03}, and hence by \eq{eq5.23} we find that
\EQ{
\wpr{\cQ_{b}}=\wpr{\cQ}\oa\wpr{\cQ}\text{  .}
}

Suppose next that we add a fresh qubit $q$ to \cQ's context, and write $\cQ_{q}$ for the resulting program.  The forward semantics of $\cQ_{q}$ is given by
\EQ{
\sem{\cQ_{q}} \left(\begin{array}{c|c}\rho_{1} & \rho_{2} \\\hline \rho_{3} & \rho_{4}\end{array}\right)
= \left(\begin{array}{c|c}\cE(\rho_{1}) & \cE(\rho_{2}) \\\hline \cE(\rho_{3}) & \cE(\rho_{4})\end{array}\right)\text{  ,}
}
which we write more concisely as
\EQ{
\sem{\cQ_{q}}=\left(\begin{array}{c|c}\cE & \cE \\\hline \cE & \cE \end{array}\right)\text{.}
}
Accordingly, we find that
\EQ{
\wpr{\cQ_{q}}=\left(\begin{array}{c|c}\wpr{\cE} &\wpr{\cE} \\\hline \wpr{\cE} & \wpr{\cE} \end{array}\right)\text{.}
}

\paragraph{Iteration.} Consider a flow chart which is obtained from a program $\cQ$ by
introducing a loop.  As explained in the above, the semantics of
the flow chart is given by the monoidal trace $\Tr{\cE}$, where \cE is the semantics of the flow chart obtained from \cQ by removing the loop.  For a predicate $M$ we have that 
\EQ{
\tr{M.(\Tr{\cE})(\rho)}=\tr{\wpr{\Tr{\cE}}(M).\rho}
\label{eq6.11}
}
By iterating explicitly and using \eqs{eq4.1}{eq5.13} we obtain

\EQ{\SP{
& \tr{M.(\Tr{\cE})(\rho)}\\
&=
\tr{M.(\cE_{11}+\sum_{i=0}^{\infty}\cE_{21};\cE_{22}^{i};\cE_{12})(\rho)}\\ 
&=\tr{M.\cE_{11}(\rho)}+\sum_{i=0}^{\infty}\tr{M.
  (\cE_{21};\cE_{22}^{i};\cE_{12})(\rho)} \\ 
&=\tr{\wpr{\cE_{11}}(M).\rho}+\sum_{i=0}^{\infty}\tr{(\wpr{\cE_{12}};
\wpr{\cE_{22}}^{i};\wpr{\cE_{21}})(M).\rho}\\ 
&=\tr{(\wpr{\cE_{11}}+
\sum_{i=0}^{\infty}\wpr{\cE_{12}};\wpr{\cE_{22}}^{i};\wpr{\cE_{21}})(M).\rho}\text{  .}
\label{eq6.12} 
}}
Comparing \eqs{eq6.11}{eq6.12}  we obtain that 
\EQ{
\wpr{\Tr{\cE}}=\wpr{\cE_{11}}+
\sum_{i=0}^{\infty}\wpr{\cE_{12}};\wpr{\cE_{22}}^{i};\wpr{\cE_{21}}\text{  .}
\label{eq6.13}
}
Moreover, the existence of the limit in \eq{eq6.13} is
guaranteed due to \pro{prop2.24}.

\begin{figure}

\scalebox{.50}{\includegraphics{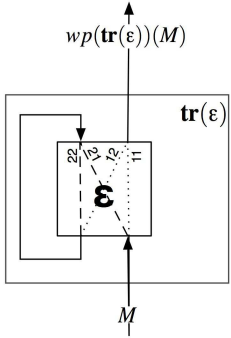}}
\caption{Iteration schematically.}

\label{iteration}
\end{figure}

\paragraph{Recursion.} Consider an operation which is defined recursively, i.e.  an operation
$\cE$ satisfying the equation $\cE=F(\cE)$, where $F$ is a flow chart.  
The required  fixed point solution to this recursive equation is given
by \eqs{eq4.2}{eq4.3}.  If we work out the weakest precondition relations
using \eq{eq4.2} and the fact that weakest precondition predicate transformers are Scott-continuous we obtain
\EQ{\SP{
\tr{M.\cE(\rho)}&= \tr{M.(\lub{i}F_{i})(\rho)} \\
&=\tr{\wpr{ \lub{i}F_{i}}(M).\rho}\\
&= \tr{(\lub{i}\wpr{F_{i}})(M).\rho} \text{  .}
\label{eq6.23}
}}
Combining this result with \pro{prop3.1} we find that the
weakest precondition predicate transformer for a recursively defined
operation $\cE=F(\cE)$ is obtained as 
\EQ{
\wpr{\cE}=\lub{i}\wpr{F_{i}}=\lub{i}\wpr{\Phi_{F}^{i}(0)}\text{  .}
\label{eq6.24}
}
The existence of the least upper bound in \eq{eq6.24} is
guaranteed by \pro{prop2.24}.  This result depends of course
on the concrete recursive specification considered.  Specifically, one
needs to determine $\Phi_{F}$ in order to determine the weakest
precondition predicate transformer corresponding to an operation $\cE$,
defined recursively as $\cE=F(\cE)$.

\section{Applications}\label{sec7}

In this section we look at some specific situations and their weakest
precondition predicate transformers.

\subsection{Grover's algorithm}

We first look into Grover's
algorithm, also known as the database search algorithm \cite{Grover96}. The algorithm is parameterized by the number of qubits $n$ and is specified in QPL as follows, where we write $N$ for $2^{n}$.
\EQ{\SP{
Grover(N)::=\: &\mbf{new \; qint_{n}}\;q:= \frac{1}{\sqrt{N}}\sum_{i=0}^{N-1}\ket{i}\; ;\\
& \mbf{new \; int} \; n:=\mbf{C}\; ;\\
& \mbf{while}\;\; \mbf{C}>0 \;\;\mbf{do} \\
&\qquad q \ast= G \; ;\\
&\qquad C:= C-1 \; ;\\
&\mbf{measure} \; q
\label{grover}
}}
Note that we assume the presence of product types of quantum integers $qint_{n}$ -- qubit registers of size $n$ -- and integers $int$, which were elaborated in~\cite{Selinger03}, and also the presence of integer operations.

The Grover operator $G$ is given by
\EQ{
G=O;IAM\text{  ,}
\label{eq7.6}}
where $O$ is a quantum oracle, which labels solutions to the search
problem, and $IAM$ is the \emph{inversion about mean}
operation, specifically $IAM=\frac{2}{N}\sum_{i,=0}^{N-1}\ketbra{i}{j}-I$.  

Supposing the solution to the search problem is given by $s$, then the relevant postcondition for Grover is given by $\bigoplus_{i=0}^{N-1}\proj{s}$, in particular we wish to obtain
\EQ{
\tr{\bigoplus_{i=0}^{N-1}\proj{s}\rho_{f_{i}}}=1\text{  ,}
\label{postg}
}
where $\bigoplus_{i=0}^{N-1}\rho_{f_{i}}$ is the final state of the algorithm, and the tuple summation is present due to measurement branching. 

We work our way backwards through the algorithm using \eq{eq5.13} in order to find the weakest precondition corresponding to the postcondition $\bigoplus_{i=0}^{N-1}\proj{s}$. First we derive the weakest precondition for the measurement in the last step of the algorithm. We do this according to a generalization of \eq{measurewp} for $N$-valued measurements, as follows.
\EQ{ \SP{
\wpr{\mbf{measure} \; q}(\bigoplus_{i=0}^{N-1}\proj{s})& = \wpr{\cE_{0}\oa\dots \oa\cE_{N-1}}(\bigoplus_{i=0}^{N-1}\proj{s})\\
&= \wpr{\cE_{0}}(\proj{s})+\dots+\wpr{\cE_{N-1}}(\proj{s})\\
&= P_{0 }\proj{s} P_{0 }+ \dots + P_{N-1 }\proj{s} P_{N-1}\\
&=\proj{s} \text{  .}
\label{measureg}
}}
Note that, since the remainder of the algorithm consists of unitary evolution, all relevant preconditions continue to be pure state projectors. 
In this case \eq{postg} holds only if the output state equals the predicate, that is if $\rho_{f}=\proj{s}$, so that pure state preconditions are at the same time the states required for the algorithm to satisfy \eq{postg} after termination.

We now focus in the while loop in the algorithm. Geometrically, the Grover operator is a rotation in the
two-dimensional space~\cite[Sec. 6.1.3]{Nielsen00}  spanned by the states $\ket{s}$ and

\EQ{
\ket{\al}=\frac{1}{\sqrt{N-1}}\sum_{x\neq s}\ket{x}
\label{eq7.71}
}More specifically, $G$ can be decomposed as 

\EQ{
G=\begin{pmatrix} \cos\theta & -\sin\theta \\ \sin\theta & \cos\theta
 \end{pmatrix} 
 \quad  \mbox{with } \sin \theta=\frac{2\sqrt{N-1}}{N}\text{  .}
 \label{eq7.73}
 }
Applying again \eq{eq5.13}, we obtain as weakest precondition with
respect to the while loop the following, 

\EQ{
\wpr{ \mbf{while}\;\; C>0 \;\;\mbf{do} \; q \ast= G}(\proj{s})=(G^{C}){\proj{s}}G^{C}\text{  ,}
\label{eq7.8}}

where we omit explicit weakest precondition reasoning for the purely classical command $C:=C-1$. Using \eq{eq7.73}, we see that $(G^{C})^{\dag}\ket{s}$
corresponds to $C$ rotations over an angle of $-\theta$ in the state
space spanned by $\ket{\al}$ and $\ket{s}$.  By choosing $C=\arccos\frac{1}{\sqrt{N}}$~\cite[Sec 6.1.3]{Nielsen00}, one rotates the postcondition $\proj{s}$ towards the precondition $\proj{\psi_{i}}$, where $\ket{\psi_{i}}$ is the initial state of the algorithm, i.e. the equal superposition state, which lies in the space spanned by the states $\ket{\al}$ and $\ket{s}$. In other words, using \eq{eq3.1} and  \eq{postg} we obtain that for all $\rho_{i}$
\EQ{\SP{
 \tr{\wpr{Grover}(\proj{s})\rho_{i}} &=\tr{\proj{s} Grover(\rho_{i})}\\
 \iff \tr{\proj{\psi_{i}}\rho_{i}}&=1 \text{  .}
}}
That is, \eq{postg} holds if and only if $\rho_{i}=\proj{\psi_{i}}$ which
is the case by construction of the algorithm. Hence we have established the
correctness of the algorithm via our backwards semantics.  

We note that an alternative derivation for Grover's algorithm based on
probabilistic weakest preconditions has been reported
in~\cite{Butler99}. However, the use of probabilistic notions only works
there because Grover's algorithm is considered for pure states only. The
mathematical structures underlying their analysis is that of probabilistic
weakest preconditions, which are in fact not suited at all for a
generalized quantum setting, as we have stressed in \sec{sec2}. In our
setting we could reason about mixed state solutions to Grover and compare
them with the pure state solution elaborated in the above. Also, while it
may seem at first sight that in~\cite{Butler99} the value of $C$ is derived
via the backward semantics this is in fact not the case. Instead, a
recurrence relation for amplitudes occurring in each Grover iteration is
solved; these amplitudes are found by applying the Grover iteration
backwards, just as we did. We chose to adhere to the interpretation of $G$
as a rotation in a two-dimensional state space in order to find $C$; we
could just as well have adhered to the derivation in~\cite{Butler99}. While
their proof is an ingenious alternative to that in~\cite{Nielsen00}, it is
not based on the theory of probabilistic weakest preconditions.

\subsection{Tossing a coin}\label{coin}
As a second application, we derive the weakest precondition for the flow
chart implementing a fair coin toss~\cite[Example 4.1]{Selinger03}. In QPL
terms the flow chart is specified as follows, where $r$ is an input qubit
register of unspecified length. 
\EQ{\SP{
coin (r) ::=\: &\mbf{new \; qbit}\;q:= \mbf{0} ;\\
&q \ast= H ;\\
&\mbf{measure} \; q;\\
&\mbf{discard} \; q
}}

An arbitrary postcondition for this program is of the form $M_{1}\oa M_{2}$, where $M_{1}$ and $M_{2}$ are both predicates over $\predv{\mbb{C}^{2n}}$ and $n$ is the number of qubits in the register $r$. We derive the corresponding weakest precondition by flowing backwards through the program, starting with the discard operation. The latter induces the following quantum operation, where $I_{N}$ is the $(N\times N)$ identity map  with $N=2^{n}$ as before, 0 denotes the $(N\times N)$ zero block matrix, and $\rho$ is a density matrix in $\den{\mbb{C}^{2(n+1)}}$,
\EQ{
 \sem{\mbf{discard} \; q}(\rho)=\left(\begin{array}{c|c}I_{N} & 0\end{array}\right)\rho\left(\begin{array}{c}I_{N} \\\hline 0\end{array}\right) + \left(\begin{array}{c|c}0 & I_{N}\end{array}\right)\rho\left(\begin{array}{c}0 \\\hline I_{N}\end{array}\right)\text{  .}
}
This leads to the following weakest precondition,
\EQ{
\wpr{\mbf{discard} \; q}(M_{1}\oa M_{2})= \left(\begin{array}{c|c}M_{1} & 0 \\\hline 0 & 0\end{array}\right)\oa \left(\begin{array}{c|c}0 & 0 \\\hline 0 & M_{2}\end{array}\right) \text{  .}
\label{discard}
}
Next, we have the measurement step. We just give the result here, as this type of derivation was already encountered in the Grover example above.
\EQ{
\wpr{\mbf{measure} \; q}[ \left(\begin{array}{c|c}M_{1} & 0 \\\hline 0 & 0\end{array}\right)\oa \left(\begin{array}{c|c}0 & 0 \\\hline 0 & M_{2}\end{array}\right)]
= \left(\begin{array}{c|c}M_{1} & 0 \\\hline 0 & M_{2}\end{array}\right)\text{  .}
\label{measurecoin}
}
The Hadamard transformation is straightforward and leads to
\EQ{
\wpr{q \ast= H} \left(\begin{array}{c|c}M_{1} & 0 \\\hline 0 & M_{2}\end{array}\right)
=  \left(\begin{array}{c|c}HM_{1}H & 0 \\\hline 0 & HM_{2}H\end{array}\right)\text{  .}
\label{hadamard}
}
Finally we move through the first command in the coin toss program, namely the addition of a new qubit. The forward semantics of this command is as follows, where $\rho$ is a density matrix in $\den{\mbb{C}^{2n}}$,
\EQ{
 \sem{\mbf{new \; qbit}\;q:= \mbf{0}}(\rho)=\left(\begin{array}{c}I_{N} \\\hline 0\end{array}\right) \rho\left(\begin{array}{c|c}I_{N} & 0\end{array}\right) \text{  .}
}
Hence, we obtain the following,
\EQ{
\wpr{\mbf{new \; qbit}\;q:= \mbf{0}} \left(\begin{array}{c|c}HM_{1}H & 0 \\\hline 0 & HM_{2}H\end{array}\right)
= HM_{1}H\text{  .}
\label{newqubit}
}

Wrapping all individual steps of the coin toss program up into one weakest precondition predicate transformation according to \eq{eq5.13} we obtain
\EQ{
\wpr{coin(r)} (M_{1}\oa M_{2})
= HM_{1}H\text{  .}
\label{coinwpr}
}

\subsection{Stabilizers are predicates}

The stabilizer formalism is an alternative description of quantum
states~\cite{Gottesman99b}.  Instead of describing states as vectors in a suitable Hilbert
space, they are described by a set of operators which leave the state
invariant.  Concretely, for an $n$-qubit system these operators are
taken from the Pauli group $G_{n}$, i.e. the group of $n$-fold tensor
products of the Pauli matrices with factors $\pm1, \pm i $ in
front.  Note that if we allow all positive operators instead one obtains the
more familiar density matrix formalism.  Of course not all states can
be described in this way.  Formally, a \emph{stabilizer state} is a
simultaneous eigenvector of an abelian subgroup of the Pauli group with
eigenvalue 1.  This subgroup is then called the \emph{stabilizer} $S$
of this state, and usually represented by its generators.  
Surprisingly, some forms of entanglement, such as graph states for
example \cite{Raussendorf02a}, as well as all Clifford group
operations, can be described efficiently via stabilizers -- a celebrated result known as the \emph{Gottesman-Knill theorem}~\cite[Sec. 10.5.4]{Nielsen00}.  This is
because for an $n$-qubit stabilizer state its stabilizer $S$ has $n-1$
generators (as opposed to $2^{n}$ amplitudes in the state formalism).  A
nice overview of stabilizer theory can be found in
\cite[Ch.~10]{Nielsen00}.

Stabilizers, which are unitaries,  fit well within
the setting of weakest preconditions, because when restricting
ourselves to pure states, they \emph{are} in fact quantum predicates.  This
follows from the following theorem.

\PRO{
Given a pure state $\rho=\proj{\psi}$ and a unitary $U$ we have that
\EQ{
\tr{U\rho}=1 \iff U\gket=\gket
\label{stabs}
}
\label{stabsprop}
}
\begin{proof}
For the left to right direction, we have that
\EQ{\SP{
&\tr{U\proj{\psi}}=\mel{\psi}{U}{\psi}=1\\
&\Rar (\bra{\psi}-\bra{\psi}U^{\dag})(\ket{\psi}-U\ket{\psi})=0\\
&\Rar \ket{\psi}-U\ket{\psi}=0\\
&\Rar \ket{\psi}=U\ket{\psi}\text{  .}
}}
The other direction is obvious.
\end{proof}

For example, consider the creation of a Bell state
$\ket{B}=\frac{\ket{00}+\ket{11}}{\sqrt{2}}$ by applying $U=CNOT.  (H
\ox I)$ to $\ket{00}$.  The stabilizer of $\ket{B}$ is generated by
$Z_{1} Z_{2}$ and $X_{1}X_{2}$.  Hence by the above result we have
$\tr{Z_{1}Z_{2}\cE_{U}(\proj{\psi})}=\tr{X_{1}X_{2}\cE_{U}(\proj{\psi})
}=1$, where $\ket{\psi}$ is the initial state of the algorithm and $\cE_{U}(\rho)=U\rho U^{\dag}$ for all $\rho$.  Applying \eq{eq3.3}, we obtain as weakest preconditions
$\wpr{\cE_{U}}(Z_{1}Z_{2})=Z_{2}$ and
$\wpr{\cE_{U}}(X_{1}X_{2})=Z_{1}$.  By \pro{prop3.1} we thus also
have $\tr{Z_{1}\proj{\psi}}=\tr{Z_{2}\proj{\psi}}=1$.  But then by the
above result $Z_{1}$ and $Z_{2}$ are stabilizers of $\gket$.  Hence
$\gket=\ket{00}$, as required.

\section{Conclusions} \label{sec8}

In this article, we have developed the predicate transformer and
weakest precondition formalism for quantum computation.  We have done
this by first noting that the quantum analogue to predicates are
expectation values of quantum measurements, given by the expression
$\tr{M\rho}$.  Then we have defined the concept of weakest
preconditions within this framework, proving that a weakest
precondition exists for arbitrary completely positive maps and
observables.  We have also worked out the weakest precondition
semantics for the Quantum Programming Language (QPL) developed in
\cite{Selinger03}.  QPL is the first model for quantum computation with
a denotational semantics, and as such the first serious attempt to
design a quantum programming language intended for programming quantum
algorithms compositionally.

With this development in place one can envisage a goal-directed
programming methodology for quantum computation.  Of course one needs
more experience with quantum programming idioms and the field is not
yet ready to produce a ``quantum'' Science of Programming.  It is likely that in the field of communication protocols, such as those based on
teleportation, we have a good stock of ideas and examples which
could be used as the basis of methodologies in this context.  

The most closely related work - apart from Selinger's work on his
programming language - is the work by Sanders and
Zuliani~\cite{Sanders00} which develops a guarded command language used
for developing quantum algorithms.  This is a very interesting paper
and works seriously towards developing a methodology for quantum
algorithms.  However, they use probability and nondeterminism to
capture probabilistic aspects of quantum algorithms.  Ours is an
\emph{intrinsically quantum} framework.  The notion of weakest
precondition that we develop here is not related to anything in their
framework.  There are other works~\cite{Baltag04} - as yet unpublished
- in which a quantum dynamic logic is being developed.  Clearly such
work will be related though they use a different notion of pairing.
Also the work in \cite{Edalat04} is related and merits further
investigation.  Edalat uses the interval domain of reals rather than the
reals as the values of the entries in his density matrices.  This seems
a good way to deal with uncertainty in the values.

There is a large literature on probabilistic predicate transformers
including several papers from the probabilistic systems group at
Oxford.  A forthcoming book~\cite{Morgan04} gives an expository account
of their work.  We emphasize again that the theory of probabilistic
predicate transformers does not capture the proper notions appropriate for
the quantum setting.  Linearity and complete positivity are essential
aspects of the theory of quantum predicate transformers.  If one tries to
work with probabilistic predicates alone one will not be able to express
healthiness conditions that capture the physically allowable
transformations, as the example presented in \sec{sec3} illustrates.

One might worry that the predicates are too restricted.  There are many
``observables'' in physics that are not positive; for example, the
$z$-component of angular momentum, written $J_z$, for a spin $\frac12$
system takes on the values $\pm\frac12$.  However, for reasoning about the
evolution of $J_z$ one can work instead with the operator $\frac12[I +
J_z]$ which has eigenvalues $\frac14$ and $\frac34$ and so is a predicate.
Of course one cannot do this for unbounded operators like the energy, but
this will not be a handicap for quantum computation.

One pleasant aspect of the present work is that it is language
independent; though we have used it to give the semantics of QPL the
weakest precondition formalism stands on its own.  We can therefore
apply it to other computational models that are appearing, for example the
one-way model~\cite{Raussendorf01,Raussendorf02a} for which
language ideas are just emerging~\cite{Danos04b}.

\section*{Acknowledgements} It is a pleasure to thank Samson Abramsky,
Bob Coecke, Elham Kashefi and Peter Selinger for helpful discussions.
Comments by the referees were very helpful.  

\bibliographystyle{apalike}

\bibliography{references}

\end{document}